%% file: main.tex
\documentclass[conference]{IEEEtran}
\IEEEoverridecommandlockouts
\usepackage{cite}
\usepackage{amsmath,amssymb,amsfonts}
\usepackage{algorithmic}
\usepackage{graphicx}
\usepackage{textcomp}
\usepackage[table,dvipsnames]{xcolor}

\usepackage{wrapfig}
\usepackage{caption}
\usepackage{algorithm}
\usepackage{algorithmic}
\usepackage{bm}
\usepackage{graphicx}
\usepackage{amsmath}
\usepackage{amssymb}
\usepackage{mathtools}
\usepackage{amsthm}
\usepackage{multirow}
\usepackage{tabularx}
\usepackage{subcaption}
\usepackage{tcolorbox}
\usepackage{enumitem}
\usepackage[formats]{listings} 
\usepackage{threeparttable}
\usepackage{lscape} 
\usepackage{booktabs}
\usepackage{amssymb}   
\usepackage{pifont}    
\newcommand{\xmark}{\ding{55}}
\usepackage{enumitem}
\usepackage{longtable}
\usepackage{makecell}
\usepackage{float}
\usepackage{caption}
\begin{document}

\title{SMILE: Speech Meta In-Context Learning for Low-Resource Language Automatic Speech Recognition}

\author{\IEEEauthorblockN{Ming-Hao Hsu}
\IEEEauthorblockA{
\textit{Electrical Engineering Department}\\
\textit{
National Taiwan University}\\
Taipei, Taiwan\\
hsuminghao1006@gmail.com}
\and
\IEEEauthorblockN{Hung-yi Lee}
\IEEEauthorblockA{
\textit{Electrical Engineering Department}\\
\textit{
National Taiwan University}\\
Taipei, Taiwan\\
hungyilee@ntu.edu.tw}
}

\maketitle

\begin{abstract}
\input{text/1_abstract}
\end{abstract}


\input{text/2_introduction}
\input{text/3_related_works}
\input{text/4_methodology}
\input{text/5_experiments}
\input{text/6_conclusion}

\bibliographystyle{abbrv}
\bibliography{refs}

\newpage
\onecolumn
\input{text/appendix}

\end{document}

%% file: text/1_abstract.tex
Automatic Speech Recognition (ASR) models demonstrate outstanding performance on high-resource languages but face significant challenges when applied to low-resource languages due to limited training data and insufficient cross-lingual generalization. 
Existing adaptation strategies, such as shallow fusion, data augmentation, and direct fine-tuning, either rely on external resources, suffer computational inefficiencies, or fail in test-time adaptation scenarios.
To address these limitations, we introduce Speech Meta In-Context LEarning (SMILE), an innovative framework that combines meta-learning with speech in-context learning (SICL). 
SMILE leverages meta-training from high-resource languages to enable robust, few-shot generalization to low-resource languages without explicit fine-tuning on the target domain. 
Extensive experiments on the ML-SUPERB benchmark show that SMILE consistently outperforms baseline methods, significantly reducing character and word error rates in training-free few-shot multilingual ASR tasks.

%% file: text/2_introduction.tex
\section{Introduction}

Automatic Speech Recognition (ASR) models, such as Whisper~\cite{radford2022robustspeechrecognitionlargescale} and SeamlessM4T~\cite{communication2023seamlessm4tmassivelymultilingual}, have transformed speech-driven applications, enabling real-time transcription and voice interaction with AI assistants. 
Despite remarkable performance in high-resource languages (HRLs) like English or Chinese, these foundation models severely underperform on low-resource languages (LRLs), primarily due to data scarcity, high linguistic variability, and inadequate cross-language generalization.

To address these limitations, adaptation strategies including shallow fusion~\cite{hu2023massivelymultilingualshallowfusion, ogawa2023iterativeshallowfusionbackward, chen2024its, li-niehues-2025-enhance, sun2023contextualbiasingremaineffective, dezuazo2025whisperlmimprovingasrmodels}, data augmentation~\cite{Park_2019, yang2024enhancinglowresourceasrversatile, venkateswaran-liu-2024-looking, bhogale24_interspeech}, and direct fine-tuning on target languages~\cite{song2024lorawhisperparameterefficientextensiblemultilingual, chen-etal-2024-meta} have gained significant attention. 
However, shallow fusion heavily relies on external language models, often lacking representations for truly low-resource languages. 
Data augmentation frequently introduces unrealistic acoustic variations, impairing generalization. 
Direct fine-tuning suffers from severe computational overhead and catastrophic forgetting when simultaneously applied to multiple languages.

Meta-learning approaches~\cite{hsu2019metalearningendtoendlowresource, xiao2021adversarialmetasamplingmultilingual} partially alleviate these issues by leveraging prior knowledge from high-resource languages for efficient adaptation to new languages.
Yet, existing meta-learning methods typically require language-specific fine-tuning and suffer from large GPU memory usage, resulting in computational and storage inefficiencies, especially for large models, and still lack true training-free few-shot generalization capabilities~\cite{Raimon_2025}.

\begin{table*}[ht]
\centering
\caption{Comparison of common ASR adaptation methods for low-resource languages.}
\label{tab:adaptation_methods}
\begin{tabular}{@{} l l l c c @{}}
\toprule
Method / Model      & Architecture & Adaptation Strategy          & Target Domain Fine-Tune & Extra Modules \\
\midrule
wav2vec 2.0 / XLS-R & Enc-only (SSL)            & Supervised fine-tuning        & \checkmark{}        & \xmark{}      \\[2pt]
Shallow Fusion      & Enc-Dec / Enc-only        & Beam-score fusion & \checkmark{}   & \checkmark{}  \\[2pt]
SICL                & Enc-Dec                   & In-context prompting          & \xmark{}            & \xmark{}      \\[2pt]
\textbf{SMILE (Ours)} & Enc-Dec                 & Meta in-context adaptation    & \xmark{}            & \xmark{}      \\
\bottomrule
\end{tabular}
\end{table*}

Conversely, Speech In-Context Learning (SICL)~\cite{10446502} leverages test-time prompts, eliminating the need for additional training data.
Similar to the right side of Fig.~\ref{fig:SMILE}, SICL leverages paired speech and text inputs to condition model decoding without further fine-tuning. 
Nevertheless, SICL performs reliably only on previously encountered dialects and significantly deteriorates with low-resource languages, severely limiting its practical applicability in test-time adaptation.

To overcome these fundamental limitations, we introduce \textbf{S}peech \textbf{M}eta \textbf{I}n-Context \textbf{LE}arning (SMILE), a framework integrating meta-training with an SICL form. 
Unlike conventional meta-learning methods that explicitly optimize initial states for new task adaptation, SMILE learns a generalized prompting strategy from high-resource languages, enabling test-time few-shot generalization to low-resource languages.
Specifically, SMILE leverages learned relationships between prompt-target pairs during meta-training, eliminating the need for explicit fine-tuning for adaptation and significantly reducing computational overhead.

Evaluated on the ML-SUPERB multilingual benchmark~\cite{shi2025mlsuperbmultilingualspeechuniversal}, SMILE consistently surpasses existing methods, demonstrating robust performance in test-time adaptation scenarios.
Ablation studies further validate the importance of the two prompt modalities, audio and text.

Our contributions are threefold:
\begin{itemize}
    \item We introduce \textbf{SMILE}, a framework that delivers robust few-shot performance for low-resource languages at test time without any explicit fine-tuning on the target domain—unlike previous few-shot methods.
    \item In contrast to conventional speech in-context learning, SMILE can adapts to out-of-domain unsupported languages using just a single prompt.
    \item Extensive evaluations on ML-SUPERB demonstrate that SMILE consistently outperforms competitive baselines, validating our design choices and effectiveness in practical low-resource ASR scenarios.
\end{itemize}

In this work, we target the test-time adaptation regime, where no low-resource language audio or text is available for training, validation, or hyperparameter tuning. 
Under this setting, only models capable of adapting purely at test time, such as those leveraging in-context prompting, are practically feasible. 
We summarize a comparison among common ASR adaptation methods in Table~\ref{tab:adaptation_methods}.
Encoder-only self-supervised learning (SSL) models like wav2vec 2.0/XLS-R possess minimal test-time adaptation ability and thus lie outside the scope of our primary evaluation, although we still provide comparative results under a few-shot fine-tuning setting in Table~\ref{tab:mlsuperb_comparison}.

%% file: text/3_related_works.tex
\section{Related Work}
\subsection{Self-Supervised Learning in Low-Resource ASR}
Self-Supervised Learning (SSL) has emerged as a dominant paradigm in ASR, primarily by learning representations from vast amounts of unlabeled speech data.
Models such as wav2vec 2.0~\cite{baevski2020wav2vec20frameworkselfsupervised}, HuBERT~\cite{hsu2021hubertselfsupervisedspeechrepresentation}, and XLS-R~\cite{conneau2020unsupervisedcrosslingualrepresentationlearning} are examples of SSL approaches that pre-train encoders on large-scale datasets to capture intricate acoustic and phonetic characteristics.
Typically, these pre-trained SSL models are adapted to downstream ASR tasks through supervised fine-tuning on labeled data specific to the target languages~\cite{anidjar2024whisperturnsstrongeraugmenting, boito2024mhubert, singh2023novel}.
Although these models can achieve strong performance, particularly in few-shot settings when fine-tuned on target low-resource languages, their test-time few-shot adaptation capabilities for low-resource languages can be minimal without some form of task-specific or language-specific tuning~\cite{lin2022listenadaptbetterwer}.
Our work, SMILE, contrasts with these traditional SSL fine-tuning approaches by aiming for enhanced data and computational efficiency, particularly in test-time adaptation scenarios, by learning to leverage prompts rather than requiring end-to-end fine-tuning.

\subsection{Meta-Learning for Low-Resource ASR}
Meta-learning seeks a language-agnostic initialization that can be adapted to new languages with only a handful of examples. Proof-of-concept studies showed that Model-Agnostic Meta-Learning (MAML) accelerates cross-lingual adaptation and outperforms multitask pre-training~\cite{hsu2019metalearningendtoendlowresource}. Subsequent work improved stability and efficiency: multi-step loss supervision~\cite{Singh_2022} and adaptive task weighting~\cite{chen-etal-2024-meta} reduce gradient variance, while parameter-efficient variants—e.g., TAMML-Light, which restricts updates to the classifier head~\cite{Chen2024Lightweight}, and adapter-based meta-initialization within SSL backbones~\cite{9414959}—retain most of the gains at a fraction of the compute.

Despite these advances, two practical hurdles remain.  
(i) All methods still require a language-specific fine-tuning phase, so test-time adaptation is almost unattainable.  
(ii) The meta-training loop performs back-to-back inner- and outer-gradient computations, causing a memory footprint that scales with model size~\cite{DBLP:journals/jmlr/WangYYY23}; training a 1.5-billion-parameter \textit{whisper-large-v2} typically exceeds the capacity of an 80 GB GPU even with gradient checkpointing.

\subsection{Speech In-Context Learning (SICL)}
Whereas meta-learning requires a brief fine-tuning stage, \emph{in-context learning} relies purely on prompts at inference.  
SICL~\cite{10446502} conducted the first systematic evaluation of Whisper’s speech-based ICL, showing that concatenating \(\langle\)speech, text\(\rangle\) exemplars before the target utterance reduces WER by 8 to 32\% on Chinese dialect and speaker adaptation tasks.  
Yet Whisper’s vanilla ICL is sensitive to prompt length and cannot easily generalize to unsupported languages.  

%% file: text/4_methodology.tex
\section{Speech Meta In-Context Learning}
\begin{figure*}
    \centering
\includegraphics[width=1\linewidth]{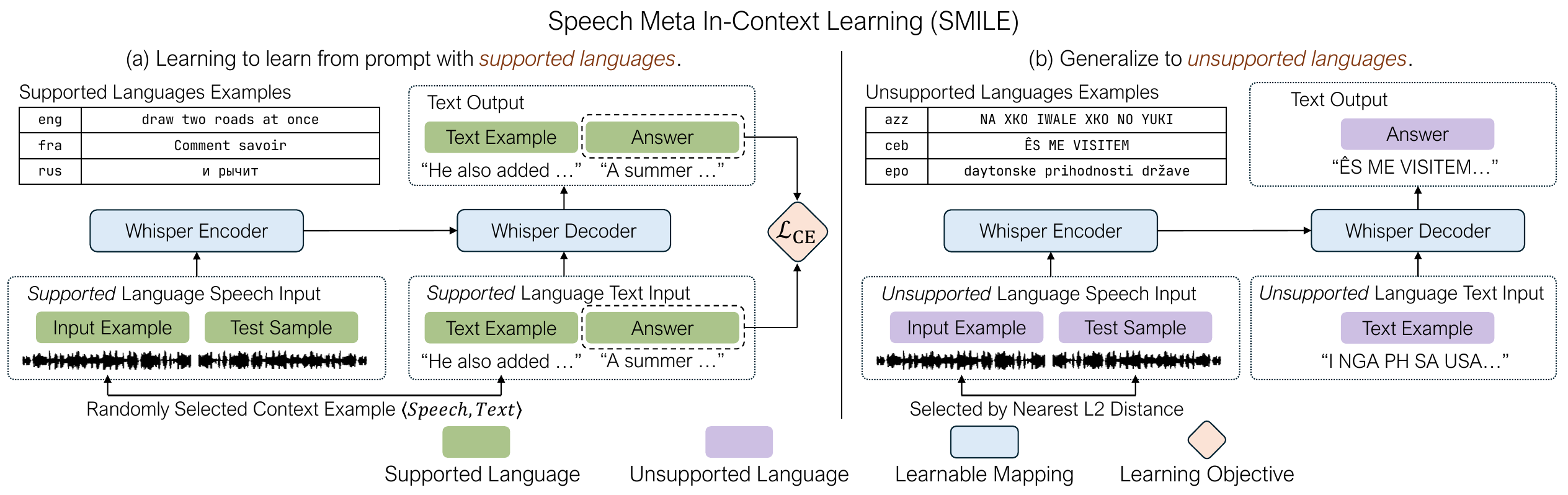}
\caption{\textbf{Speech Meta In-Context Learning (SMILE).}  
SMILE is meta-trained on high-resource languages so the model can exploit prompts without any fine-tuning.  
\textbf{(a) Meta-training—learning to learn from prompts:} For each step, we randomly sample prompt–target pairs from high-resource languages, mask the prompt tokens, and compute the cross-entropy loss $\mathcal{L}_{\mathrm{CE}}$ only on the target tokens, fostering language-agnostic generalization.  
\textbf{(b) Inference-time adaptation:} At inference time, we retrieve the prompt whose encoder embedding has the smallest Euclidean (L2) distance to the target utterance, concatenate it with the utterance, and decode the transcription.}
    \label{fig:SMILE}
\end{figure*}

Our approach, \textbf{Speech Meta In-Context Learning (SMILE)}, leverages the prompt-based in-context learning mechanism from high-resource languages to enhance model generalization to unsupported, low-resource languages. 
Specifically, we meta-fine-tune Whisper to utilize a pair of prompt-target example, allowing the model to generalize to novel languages without explicit fine-tuning on target-language data.

\subsection{Problem Formulation}

Automatic Speech Recognition (ASR) systems aim to transform input speech waveforms $\boldsymbol{x}$ into corresponding textual transcriptions $\boldsymbol{y}$.
Formally, the ASR task is to learn a probabilistic mapping \(p_\theta\) parameterized by \(\theta\), modeling the conditional probability:
\[
    \hat{\boldsymbol{y}} = \arg\max_{\boldsymbol{y}} \log p_\theta(\boldsymbol{y} \mid \boldsymbol{x}),
\]
where \(\boldsymbol{x}\) denotes the input acoustic signals and \(\boldsymbol{y}\) denotes the output text tokens. This mapping is typically optimized by minimizing a negative log-likelihood loss over annotated speech-text pairs.

In our SMILE approach, we assume access to two pieces of information: a prompt audio waveform $\boldsymbol{x}^{(\text{prm})}$ paired with its transcription $\boldsymbol{y}^{(\text{prm})}$, and a target audio waveform $\boldsymbol{x}^{(\text{tgt})}$ whose transcription $\boldsymbol{y}^{(\text{tgt})}$, we want to predict.
During training, we feed all three elements—$\boldsymbol{x}^{(\text{prm})}$, $\boldsymbol{y}^{(\text{prm})}$, and $\boldsymbol{x}^{(\text{tgt})}$ into our model, and optimize it to maximize the likelihood of the correct target transcription. 
Formally, our objective is:

\[
    \hat{\boldsymbol{y}}^{(\text{tgt})} = \arg\max_{\boldsymbol{y}^{(\text{tgt})}} \log p_\theta(\boldsymbol{y}^{(\text{tgt})} \mid \boldsymbol{x}^{(\text{prm})}, \boldsymbol{y}^{(\text{prm})}, \boldsymbol{x}^{(\text{tgt})})
\]

\subsection{Model Architecture}

SMILE is based on Whisper's encoder-decoder structure. During training, the input is constructed by concatenating the prompt audio $\boldsymbol{x}^{(\text{prm})}$ and target audio $\boldsymbol{x}^{(\text{tgt})}$:

$$
\boldsymbol{x}^{(\text{input})} = [\boldsymbol{x}^{(\text{prm})}; \boldsymbol{x}^{(\text{tgt})}]
$$

The encoder maps this concatenated waveform into latent features $\boldsymbol{h}$:

$$
\boldsymbol{h} = \mathrm{Encoder}(\boldsymbol{x}^{(\text{input})})
$$

Meanwhile, the decoder input tokens are constructed as a concatenation of special tokens (language/task indicators) and prompt text tokens:

$$
\boldsymbol{z}^{(\text{input})} = [\boldsymbol{z}_{\text{special}}; \boldsymbol{z}^{(\text{prm})}; \boldsymbol{z}^{(\text{tgt})}_{<t}]
$$

Here, $\boldsymbol{z}^{(\text{tgt})}_{<t}$ means the prefix of the target token sequence up to (but not including) timestep $t$.
The Whisper decoder generates the output tokens autoregressively:

$$
p_\theta(\boldsymbol{y}_t^{(\text{tgt})} \mid \boldsymbol{y}_{<t}^{(\text{tgt})}, \boldsymbol{x}^{(\text{input})}, \boldsymbol{y}^{(\text{prm})}) = \mathrm{Decoder}(\boldsymbol{z}^{(\text{input})}, \boldsymbol{h})
$$

\subsection{Meta-training on Supported Languages}
During every meta-training step, we \emph{exclusively} sample a
\textit{prompt-target pair} $(\boldsymbol{x}^{(\mathrm{prm})}, \boldsymbol{y}^{(\mathrm{prm})},
\boldsymbol{x}^{(\mathrm{tgt})}, \boldsymbol{y}^{(\mathrm{tgt})})$ from a language
$\ell$ from dataset $\mathcal{D}_{\text{SL}}$ construct from supported language set $S_{\mathrm{SL}}$.
No utterance from the unsupported language set $S_{\mathrm{UL}}$ is used at this stage.

To enforce effective in-context learning from prompts, we propose a special loss calculation. 
Specifically, during training, we mask out the loss contributions from prompt tokens, allowing the model to explicitly focus on predicting only the target tokens.
Therefore, the modified training loss becomes:

$$
\mathcal{L}_{\text{SMILE}} = - \log p_\theta(\boldsymbol{y}_t^{(\text{tgt})} \mid \boldsymbol{y}_{<t}^{(\text{tgt})}, \boldsymbol{x}^{(\text{input})}, \boldsymbol{y}^{(\text{prm})}),
$$
where
$$
(
\boldsymbol{x}^{(\text{prm})},
\boldsymbol{y}^{(\text{prm})},
\boldsymbol{x}^{(\text{tgt})},
\boldsymbol{y}^{(\text{tgt})}
)\sim \mathcal{D}_{\mathrm{SL}}
$$

This ensures that the gradient update during back-propagation occurs based on predictions of target tokens.

\subsection{Test-time Adaptation on Unsupported Languages}
At test time, the model parameters $\theta$ are frozen.
We evaluate SMILE and other baselines under \textit{both} (i) supported languages and (ii) unsupported languages.
During inference, we retrieve a single prompt that is acoustically closest to the test utterance.  
Let $\{\boldsymbol{x}^{(\text{prm})}_i\}_{i=1}^{N}$ denote the pool of candidate prompt waveforms.  
Each candidate and the target utterance are first passed through SMILE’s frozen encoder:

\[
\boldsymbol{h}^{(\text{prm})}_i=\operatorname{Encoder}(\boldsymbol{x}^{(\text{prm})}_i), \qquad
\boldsymbol{h}^{(\text{tgt})}=\operatorname{Encoder}(\boldsymbol{x}^{(\text{tgt})}).
\]

We then apply mean-pooling over the time axis to obtain utterance-level embeddings  
$\bar{\boldsymbol{h}}^{(\text{prm})}_i, \bar{\boldsymbol{h}}^{(\text{tgt})} \in \mathbb{R}^d$.  
The prompt whose embedding is nearest (in Euclidean distance) to the target embedding is selected:

\[
i^{\ast}=\arg\min_{i}\,\bigl\|\bar{\boldsymbol{h}}^{(\text{tgt})}-\bar{\boldsymbol{h}}^{(\text{prm})}_i\bigr\|_{2}
\]

The corresponding waveform $\boldsymbol{x}^{(\text{prm})}_{i^{\ast}}$ is concatenated with the test utterance to supply a context that is maximally aligned in the encoder feature space.

To satisfy Whisper’s 30-second input limit and reduce cross-lingual length mismatch, we limit the prompt pool to utterances shorter than 15 seconds; the same constraint is applied during training (see Section~\ref{sec:hyperparameters} for details).

%% file: text/5_experiments.tex
\section{Experimental Setup}
\subsection{Dataset}
\input{tables/main}
\label{sec:dataset}
All experiments are conducted on the \textbf{ML-SUPERB} benchmark~\cite{shi2025mlsuperbmultilingualspeechuniversal}, a corpus that covers 143 languages.
Following the official ML-SUPERB protocol, we use two ten-minute splits for fine-tuning and evaluation:

\begin{itemize}
\item \texttt{10\_min\_train}: ten minutes \textit{per language} for fine-tuning (37.43 hours in total).
\item \texttt{10\_min\_test}: ten minutes \textit{per language} for evaluation and prompt-pool construction (44.97 hours in total).
\end{itemize}

Among the 143 languages, 89 are included in Whisper’s original pre-training corpus. 
We refer to them as \textit{high-resource} or \textit{supported} languages (SL).
The remaining 54 languages lack a dedicated Whisper language token; we therefore treat them as \textit{low-resource}, denoted \textit{unsupported} languages (UL).

For ICL-related experiments, the prompt pool is built from the \texttt{10\_min\_test} split. 
To prevent information leakage, we exclude the target utterance itself from the pool, i.e., a leave-one-out strategy at the utterance level. 
Note that we follow ML-SUPERB to map \texttt{jpn\_org} as \texttt{jpn}, treating it as a supported language.

\subsection{Baselines}
We benchmark \textbf{SMILE} against the following baselines:
\begin{itemize}
    \item \textbf{Whisper (Vanilla)}~\cite{radford2022robustspeechrecognitionlargescale}: the whisper-large-v2 model released by OpenAI, trained on 99 high-resource languages with no additional tuning.
    \item \textbf{Speech In-Context Learning (SICL)}~\cite{10446502}: performs zero-shot inference by prepending $k$ demonstration pairs at test time, exploiting Whisper’s innate ICL ability without further training.
    \item \textbf{Shallow Fusion}~\cite{hu2023massivelymultilingualshallowfusion, ogawa2023iterativeshallowfusionbackward, chen2024its, li-niehues-2025-enhance}: linear combination of acoustic and external LM scores at decoding. We follow prior work and fuse a GPT-2 language model, keeping the total parameter budget comparable across methods.
    \item \textbf{Model-Agnostic Meta-Learning (MAML)}~\cite{hsu2019metalearningendtoendlowresource}: meta-learns an initialization $\theta^{*}$ and adapts to each target language during a short inner loop. Because of GPU resource constraints, we use whisper-base and first-order MAML (FOMAML) for this baseline.
    \item \textbf{LoRA-Whisper} (LoRA/AdaLoRA)~\cite{song2024lorawhisperparameterefficientextensiblemultilingual}: parameter-efficient fine-tuning on unsupported languages via LoRA/AdaLoRA.
    \item \textbf{Whisper-LM}~\cite{dezuazo2025whisperlmimprovingasrmodels}: Whisper fine-tuned on the unsupported language with LoRA, followed by shallow-fusion with GPT-2.
\end{itemize}

\noindent We also compare the following SMILE variants for ablation:
\begin{itemize}
    \item \textbf{SMILE-UL}: trains on all unsupported languages instead of zero-shot transfer, illustrating the benefit of explicit few-shot unsupported language supervision.
    \item \textbf{SMILE-Rand-$k$}: randomly selects $k$ supported languages ($k\!<\!89$) as the meta-train pool to examine data-efficiency (see Fig.~\ref{fig:random_number}).
\end{itemize}

\subsection{Training and Evaluation Details}
\label{sec:hyperparameters}

\noindent\textbf{Data pre-processing.}
Whisper-large-v2 ingests at most 30 seconds of audio and emits up to 448 tokens per forward pass.
To make sure the prompt plus target audio will not over Whisper's 30-second window and 448 maximum output token length, we retain utterances shorter than 15 seconds and whose reference transcripts contain fewer than 220 tokens—criteria that cover 22,031 of the original 24,256 samples (90.81\%).

We employ AdaLoRA~\cite{zhang2023adaloraadaptivebudgetallocation} for efficiently fine-tuning SMILE, which adapts the rank per layer during training according to layer importance, reallocating the parameter budget to the layers that benefit most.
We target all attention projections (\texttt{q\_proj}, \texttt{k\_proj}, \texttt{v\_proj}, \texttt{out\_proj}) and feed-forward layers (\texttt{fc1}, \texttt{fc2}) in both encoder and decoder while freezing the remaining weights.
Hyper-parameters follow our codebase: \(\textit{init-r}=12 \rightarrow \textit{target-r}=4\), \(\alpha=32\), dropout \(=0.1\), \(\beta_1=\beta_2=0.85\).

\textbf{Optimization.}
All models are fine-tuned for a maximum of 300 optimization steps on a single H100 80 GB GPU.  
We use AdamW with \(\beta_1{=}0.9\), \(\beta_2{=}0.98\), and weight decay 0.01.  
The learning rate linearly warms up for 100 steps to a peak of \(1\times10^{-3}\) and then decays linearly to zero.  

\textbf{Evaluation.}
Consistent with ML-SUPERB, we report the macro-average CER and WER of all languages.
To prevent extreme outliers from obscuring cross-language trends, we remove the three worst languages.  
We analyze these cases in Section~\ref{sec:error_analysis} in detail.

\section{Results and Analysis}
\subsection{Main Performance Comparison}

Table~\ref{tab:cer-wer-comparison} reports macro-averaged character error rate (CER) and word error rate (WER) on supported languages (SL) and unsupported languages (UL).  
For methods that require learning, such as LoRA-Whisper or \textsc{SMILE} - we show the \emph{mean ± standard deviation} over three random seeds; inference-only baselines (e.g.\ SICL, Shallow-Fusion) are deterministic and therefore appear without deviations.

\textbf{Test-time few-shot adaptation.}  
SMILE is the most effective few-shot system on both language groups.
Relative to SICL, it trims an additional 15.95\% CER on unsupported languages while retaining the best-supported languages' performance, indicating that the meta-initialization has internalized the "\textit{learn-from-prompt}" behavior and can transfer it from supported languages to unsupported languages.  
Shallow-Fusion helps Whisper on supported languages because the external GPT-2 LM is well matched to some of those languages, yet it fails to assist on unsupported languages -- language models trained without unsupported languages provide little linguistic support, and decoding speed is hindered by second-pass rescoring.

\textbf{Few-shot fine-tuning.}  
With a small amount of unsupported language speech for adaptation, LoRA-Whisper improves on unsupported languages but suffers catastrophic forgetting on supported languages, performing worse than Vanilla.  
SMILE-UL, in contrast, achieves a lower CER on unsupported languages compared to most baselines while preserving CER on supported languages, confirming that its meta-training on unsupported languages enhances the model's "\textit{learn-from-prompt}" ability without erasing high-resource knowledge.
As for MAML, although target-domain fine-tuning lowers the CER on unsupported languages, it still suffers from catastrophic forgetting on supported languages to an even greater extent than LoRA-Whisper.

\textbf{Inference efficiency and GPU-memory overhead.}
We evaluate runtime cost using evaluation throughput (samples per second, $\uparrow$).
SMILE sustains 1.47 samples/sec—on par with Vanilla Whisper—because it adds no auxiliary language model. By contrast, Shallow Fusion and Whisper-LM achieve only 0.17 samples/sec, taking 8.65 $\times$ longer to process a single sample on the same GPU with batch size 1.

On the other hand, MAML training is highly memory-intensive, as it must retain activations and gradients at every inner-loop step. Consequently, with the same GPU resources we can fine-tune only the whisper-base model, which substantially limits the approach’s potential.

\subsection{Data Efficiency}

\begin{figure}
    \centering    \includegraphics[width=1\linewidth]{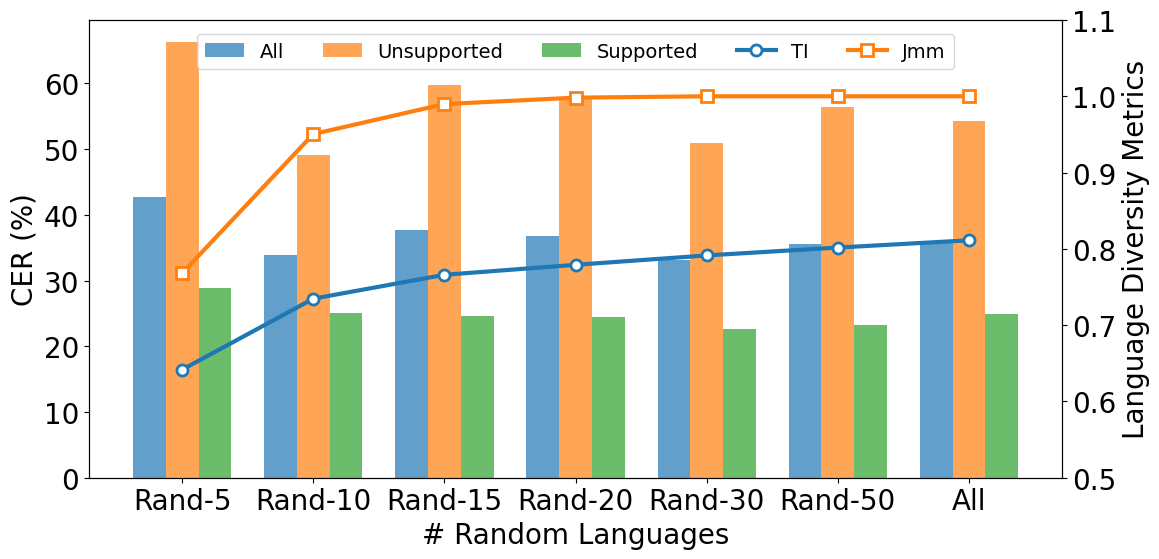}
    \caption{Mean CER (bars) for all, supported, and unsupported languages versus the number of randomly selected training languages. Lines give diversity scores: typology index (TI) and Jaccard min-max (Jmm). Performance and diversity plateau once approximately 10 random languages are included.}
    \label{fig:random_number}
\end{figure}
Fig.~\ref{fig:random_number} shows the results of SMILE-Rand-$k$ that randomly selects $k$ different languages from the supported languages for training SMILE.
This is to check how many languages we need to get similar results with all language settings.
The results demonstrate that SMILE achieves competitive performance even with a number of languages $\geq10$, confirming its data efficiency.

To further substantiate this finding, we conducted a deeper linguistic diversity analysis.
We hypothesize that SMILE benefits mainly from \emph{diversity} rather than the brute-force quantity of high-resource languages.  
To verify this, we measure the coverage of any sampled language set $S$ with two complementary metrics:

\begin{enumerate}
\item \textbf{Typology Index (TI)}~\cite{ponti2020xcopa}: This metric is the average binary entropy computed across all discretized acoustic features within the language subset $S$. A TI value of 1 indicates maximal diversity, signifying a perfectly balanced presence and absence of these features.
\item \textbf{Jaccard min-max (Jmm)}~\cite{samardzic2024jmm}: This metric calculates the Jaccard similarity between the set of discretized acoustic features present in the subset $S$ and those present in the full language set $F$. A Jmm of 1 means the subset $S$ covers all features found in the full set.
\end{enumerate}

Fig.~\ref{fig:random_number} (right axis) plots TI and Jmm versus the number of training languages. 
Both diversity metrics plateau at around 10 languages, coinciding with the observed saturation point in CER, supporting our hypothesis that phonetic diversity is more critical than sheer language quantity.
This supports our claim that approximately 10 randomly chosen languages provide sufficient phonetic coverage for the meta-training objective.

\subsection{Comparison to SSL Baselines}
\label{sec:ssl_models}
In few-shot scenarios, SSL encoders are typically fully fine-tuned, updating roughly 300 M parameters across thousands of optimization steps.
While incorporating more labeled data can boost their accuracy, it does so at a steep computational cost.
As shown in Table~\ref{tab:mlsuperb_comparison}, SMILE updates only 1.38\% of Whisper’s parameters and converges within 300 steps, yet surpasses all SSL baselines in both test-time adaptation (SMILE) and few-shot learning (SMILE-UL) settings.
These findings indicate that learning \emph{how to leverage a single prompt} is more data- and compute-efficient than pre-training a generic acoustic encoder and subsequently fine-tuning it end-to-end.
Consequently, SMILE delivers competitive or superior recognition accuracy at a fraction of the computational budget.

\subsection{Prompt Sample Selection}
We investigate several prompt sample selection strategies for choosing prompt examples in SMILE, including cosine similarity, L2 distance based on encoder features, the shortest sample based on tokenizer output, and random sampling.

Table~\ref{tab:sample-selection} presents CER and WER of all languages, demonstrating that selecting prompts based on the L2 distance of the encoder output and random sampling consistently yields the lowest CER.
However, other methods, though, cannot achieve the lowest; they still present comparable performance.
The results show the low sensitivity of prompt sample selection, showing the robustness of SMILE.

\begin{table}[t]
  \centering
  \renewcommand{\arraystretch}{1.2}
  \caption{Performance comparison to SSL model baselines on ML-SUPERB few-shot learning set.}
    \label{tab:mlsuperb_comparison}
    \setlength{\tabcolsep}{6pt}
    \begin{tabular}{@{}l c@{}}
      \toprule
      Method     & CER $\downarrow$ \\ 
      \midrule
      SMILE-UL  & 31.99\% \scriptsize $\pm$ 1.20\% \\ 
      SMILE & 35.83\% \scriptsize $\pm$ 1.06\% \\
      SICL~\cite{10446502}  & 42.31\% \\ 
      XLSR-53\textsuperscript{\textdagger}~\cite{conneau2020unsupervisedcrosslingualrepresentationlearning} & 43.60\% \\ 
      XLSR-128\textsuperscript{\textdagger}~\cite{babu2021xlsrselfsupervisedcrosslingualspeech}            & 40.90\% \\ 
  HuBERT-large\textsuperscript{\textdagger}~\cite{hsu2021hubertselfsupervisedspeechrepresentation}        & 48.20\% \\ 
      wav2vec2-large\textsuperscript{\textdagger}~\cite{baevski2020wav2vec20frameworkselfsupervised}      & 45.80\% \\ 
      \bottomrule
    \end{tabular}
  \begin{tablenotes}
      \footnotesize
      \item\textdagger Results from ML-SUPERB~\cite{shi2025mlsuperbmultilingualspeechuniversal}.
    \end{tablenotes}
\end{table}

\begin{table}[htbp]
  \centering
  \renewcommand{\arraystretch}{1.2}
  \setlength{\tabcolsep}{8pt}
    \caption{CER and WER on all languages with different prompt selection methods. Different sample selection methods would not cause a large performance gap, verifying SMILE's robustness.}
  \begin{tabular}{lcc}
    \toprule
     &
    CER $\downarrow$ &
    WER $\downarrow$ \\
    \midrule
    L2 Distance &
      35.83\% {\scriptsize$\pm$ 1.06\%} &
      65.75\% {\scriptsize$\pm$ 0.50\%} \\
    Cosine Similarity &
      38.03\% {\scriptsize$\pm$ 2.46\%} &
      65.98\% {\scriptsize$\pm$ 0.21\%} \\
    Random &
      35.77\% {\scriptsize$\pm$ 1.66\%} &
      67.15\% {\scriptsize$\pm$ 1.50\%} \\
    Token Length &
      38.86\% {\scriptsize$\pm$ 2.95\%} &
      68.56\% {\scriptsize$\pm$ 1.22\%} \\
    \bottomrule
  \end{tabular}
  \label{tab:sample-selection}
\end{table}

\subsection{Ablation Study}
To verify the contributions of \emph{prompt audio} and \emph{prompt text}, we individually remove each modality and evaluate again.
Table~\ref{tab:ablation} shows that discarding \textbf{either} component hurts performance on both supported (SL) and unsupported (UL) languages, indicating that Whisper relies on cross-modal alignment for optimal decoding.

The effect is especially pronounced for \textit{w/o Prompt Text}: its CER on unsupported languages increases by 20.83\% compared with the full model.
Our case study reveals that outputs for non-Latin unsupported languages contain $>$ 70\% Latin letters, which error pattern is absent in the \textit{Full Model}.
This suggests that text prompts implicitly inject lexical priors, steering the decoder toward valid words, correct scripts, and shorter hypotheses; without these textual cues, the model loses that guidance.

\begin{table}[t]
  \centering
  \renewcommand{\arraystretch}{1.2}
  \setlength{\tabcolsep}{8pt}
    \caption{Ablation results. We assess each modality’s contribution by removing it and reporting the resulting CER on supported (SL) and unsupported (UL) languages with a standard deviation across 3 random seeds.}
  \begin{tabular}{lcc}
    \toprule
     &
    SL CER $\downarrow$ &
    UL CER $\downarrow$ \\
    \midrule
    \textit{Full Model} & 
        24.93\% \scriptsize $\pm$ 2.46\% & 
        54.30\% \scriptsize $\pm$ 2.40\% \\
    \textit{w/o Prompt Audio} & 
        29.97\% \scriptsize $\pm$ 3.52\% &
        54.74\% \scriptsize $\pm$ 1.60\% \\
    \textit{w/o Prompt Text} & 
        29.40\% \scriptsize $\pm$ 0.19\% & 
        75.13\% \scriptsize $\pm$ 2.64\% \\
    \bottomrule
  \end{tabular}
  \label{tab:ablation}
\end{table}

\subsection{Error Analysis}
\label{sec:error_analysis}
We find that SMILE breaks down in some low-resource languages, especially \texttt{kat}, \texttt{sin}, \texttt{div}, and \texttt{nor}. 
On long utterances, the beam decoder slips into the classic repetition loop of autoregressive models, endlessly emitting a few high-probability tokens and inflating the hypothesis length~\cite{holtzman2020curiouscaseneuraltext}.
Especially, because of the existing problems, \texttt{nor} will be excluded from the ML-SUPERB benchmark evaluation~\cite{shi2025mlsuperbmultilingualspeechuniversal}.

%% file: tables/main.tex
\begin{table*}[ht]
  \centering
\renewcommand{\arraystretch}{1.2}
  \caption{Character Error Rate (CER) and Word Error Rate (WER) of various models on supported (SL) and unsupported (UL) languages (in \%). Values with standard deviations were obtained over three random seeds; values without are from deterministic methods without additional training. Methods below \textbf{Unsupported Language Fine-tuning} means the methods see the unsupported languages during the training stage.}
  \label{tab:cer-wer-comparison}
  \begin{tabularx}{\textwidth}{l *{4}{>{\centering\arraybackslash}X}}
    \toprule
    Method & SL CER $\downarrow$ & SL WER $\downarrow$ & UL CER $\downarrow$ & UL WER $\downarrow$ \\
    \midrule    Vanilla~\cite{radford2022robustspeechrecognitionlargescale} &
      37.38\% &
      65.34\% &
      86.55\% &
      127.61\% \\
    SICL~\cite{10446502} &
      25.81\% &
      51.12\% &
      70.25\% &
      99.91\% \\
    Shallow-Fusion~\cite{sun2023contextualbiasingremaineffective} &
      93.52\% & 
      127.94\% &
      225.57\% &
      282.21\% \\
    SMILE (Ours) &
      24.93\% \scriptsize $\pm$ 2.46\% &
      53.18\% \scriptsize $\pm$ 0.70\% &
      54.30\% \scriptsize $\pm$ 2.40\% &
      87.74\% \scriptsize $\pm$ 0.45\% \\
    \midrule
    \multicolumn{5}{l}{\textbf{Unsupported Language Fine-tuning}} \\
    Lora-Whisper (AdaLoRA)~\cite{song2024lorawhisperparameterefficientextensiblemultilingual} &
      42.61\% \scriptsize $\pm$ 1.77\% &
      73.56\% \scriptsize $\pm$ 1.16\% &
      44.12\% \scriptsize $\pm$ 0.51\% &
      96.66\% \scriptsize $\pm$ 1.11\% \\
    Lora-Whisper (LoRA)~\cite{song2024lorawhisperparameterefficientextensiblemultilingual} &
      41.81\% \scriptsize $\pm$ 1.94\% & 
      73.41\% \scriptsize $\pm$ 0.41\% &
      46.53\% \scriptsize $\pm$ 2.01\% &
      95.15\% \scriptsize $\pm$ 2.03\% \\
    Whisper-LM~\cite{dezuazo2025whisperlmimprovingasrmodels} &
      66.40\% \scriptsize $\pm$ 6.35\% & 
      82.21\% \scriptsize $\pm$ 1.43\% &
      77.96\% \scriptsize $\pm$ 7.67\% &
      103.30\% \scriptsize $\pm$ 0.36\% \\
    MAML\textsuperscript{\textdagger}~\cite{hsu2019metalearningendtoendlowresource} &
      63.83\% \scriptsize $\pm$ 0.78\% &
      101.09\% \scriptsize $\pm$ 4.94\% & 
      73.24\%  \scriptsize $\pm$ 3.88\%&
      112.91\% \scriptsize $\pm$ 3.41\%\\
    SMILE-UL (Ours) &
      24.69\% \scriptsize $\pm$ 0.29\% &
      56.92\% \scriptsize $\pm$ 0.49\% &
      44.37\% \scriptsize $\pm$ 2.79\% &
      81.98\% \scriptsize $\pm$ 1.44\% \\
    \bottomrule
  \end{tabularx}
  \begin{tablenotes}
      \footnotesize
      \item\textdagger Due to GPU memory limitations, we used whisper-base and first-order MAML (FOMAML) in our implementation.
    \end{tablenotes}
\end{table*}

%% file: text/6_conclusion.tex
\section{Conclusion}
We proposed SMILE, a novel integration of meta-learning and in-context learning designed to enhance test-time few-shot generalization of multilingual ASR systems.
Unlike traditional approaches, SMILE efficiently utilizes prompt-target pairs learned from high-resource languages to generalize effectively to unsupported low-resource languages at test time.
Our method significantly reduces computational overhead and storage requirements compared to conventional fine-tuning, meta-learning, and external language model strategies.
Comprehensive evaluations on the ML-SUPERB benchmark demonstrate that SMILE substantially surpasses existing baselines under similar training and inference settings.

%% file: text/appendix.tex
\appendix
\subsection{Algorithm}

We report a detailed algorithm with the pseudo code in Algorithm~\ref{alg:smile}.
Our proposed SMILE is composed of 2 parts. (i) \textbf{Meta-training phase:} For each step, we randomly sample prompt–target pairs from high-resource languages, mask the prompt tokens, and compute the cross-entropy loss $\mathcal{L}_{\mathrm{CE}}$ only on the target tokens, fostering language-agnostic generalization.  
\textbf{(ii) Inference phase:} At test time, we retrieve the prompt whose encoder embedding has the smallest Euclidean (L2) distance to the target utterance, concatenate it with the utterance, and decode the transcription.

\begin{algorithm}[H]
\caption{Speech Meta In-Context Learning (SMILE)} 
\label{alg:smile}
\textbf{Input}: High-resource language dataset $\mathcal{D}_{\text{HR}} = \{(x_i, y_i)\}$, Low-resource language test set $\mathcal{D}_{\text{LR}}$ \\
\textbf{Output}: Fine-tuned ASR model $f_{\theta^*}$ capable of zero-shot adaptation.
\begin{algorithmic}[1]
\STATE {\color{gray} // Meta-training phase: Learning to learn from prompts}
\STATE Initialize Whisper model parameters $\theta$ with pre-trained weights
\STATE Initialize AdaLoRA parameters with target budget $B_{\text{target}}$
\FOR{each meta-training iteration}
    \STATE Sample a batch of high-resource languages from $\mathcal{D}_{\text{HR}}$
    \FOR{each sample $(x, y)$ in batch}
        \STATE Randomly select prompt audio-text pair $(x^{(p)}, y^{(p)})$ from same language
        \STATE Construct input: $x^{(input)} = [x^{(p)}; x^{(t)}]$ where $x^{(t)} = x$
        \STATE Construct decoder input: $z^{(input)} = [z_{special}; z^{(p)}; z^{(t)}_{<t}]$
        \STATE Forward pass: $h = \text{Encoder}(x^{(input)})$
        \STATE Generate predictions: $p_\theta(y^{(t)}_t | y^{(t)}_{<t}, x^{(input)}, y^{(p)})$
        \STATE Compute SMILE loss: $\mathcal{L}_{\text{SMILE}} = -\sum_t \log p_\theta(y^{(t)}_t | y^{(t)}_{<t}, x^{(input)}, y^{(p)})$
    \ENDFOR
    \STATE Update $\theta$ using AdaLoRA with learning rate schedule
\ENDFOR
\STATE {\color{gray} // Inference phase: Zero-shot adaptation to low-resource languages}
\STATE Build prompt pool $\mathcal{P}$ from $\mathcal{D}_{\text{HR}}$ (utterances $< 15$ seconds)
\FOR{each test utterance $(x^{(test)}, y^{(test)})$ in $\mathcal{D}_{\text{LR}}$}
    \STATE Encode test utterance: $\bar{h}^{(t)} = \text{MeanPool}(\text{Encoder}(x^{(test)}))$
    \FOR{each prompt candidate $(x^{(p_i)}, y^{(p_i)}) \in \mathcal{P}$}
        \STATE Encode prompt: $\bar{h}^{(p_i)} = \text{MeanPool}(\text{Encoder}(x^{(p_i)}))$
        \STATE Compute distance: $d_i = \|\bar{h}^{(t)} - \bar{h}^{(p_i)}\|_2$
    \ENDFOR
    \STATE Select nearest prompt: $i^* = \arg\min_i d_i$
    \STATE Construct input: $x^{(input)} = [x^{(p_{i^*})}; x^{(test)}]$
    \STATE Construct decoder input: $z^{(input)} = [z_{special}; z^{(p_{i^*})}; z^{(test)}_{<t}]$
    \STATE Generate prediction: $\hat{y}^{(test)} = f_{\theta^*}(x^{(input)}, z^{(input)})$
\ENDFOR
\STATE \textbf{return} $f_{\theta^*}$
\end{algorithmic}
\end{algorithm}

\subsection{Hyperparameters and Baseline Implementation Details}

\textbf{Shallow-Fusion.}
We implement \textit{Shallow-Fusion} that linearly interpolates acoustic probabilities from Whisper with prior probabilities from an external language model (LM).
Concretely, we keep the \texttt{openai/whisper-large-v2} weights frozen and couple them with a frozen GPT-2 (\texttt{openai-community/gpt2}, 124 M parameters) through the \texttt{whisper-lm-transformers} pipeline:
\[
\log P(\mathbf{y}\,|\,\mathbf{x}) \;=\; \log P_{\text{ASR}}(\mathbf{y}\,|\,\mathbf{x})
\;+\; \alpha\,\log P_{\text{LM}}(\mathbf{y})
\;+\; \beta\,|\mathbf{y}|,
\]
where $\alpha=2.7$ controls LM influence and $\beta=0.0018$ serves as a length penalty.
We clear \texttt{suppress\_tokens} and \texttt{forced\_decoder\_ids} in Whisper’s \texttt{generation\_config} so that the LM may freely rescore every sub-word token.
During evaluation we employ beam search with a conservative cap of \texttt{max\_new\_tokens = 180}. 
If GPT-2’s 1024-token window is exceeded, we automatically retry with a shorter cap or truncate the hypothesis to 1000 tokens.
Ground-truth sentences longer than 180 Whisper tokens or 800 GPT-2 tokens are skipped to avoid memory overflow.
We test on the ML-SUPERB \texttt{10\_min\_test} split.

\textbf{Whisper-LM.}
This variant differs from plain shallow fusion only in that the acoustic model is \emph{fine-tuned} before decoding.
We adapt \texttt{openai/whisper-large-v2} on the ML-SUPERB \texttt{10\_min\_train} split with AdaLoRA, updating only the low-rank adaptation matrices and leaving the backbone weights frozen.  
During inference we reuse the fusion decoder, beam configuration, LM weight $\alpha=2.7$, length penalty $\beta=0.0018$, and all safety checks described for the previous baseline; the LM itself remains the frozen GPT-2.
Thus the combined score
\[
\log P(\mathbf{y}\,|\,\mathbf{x})=\log P_{\text{ASR}^{\ast}}(\mathbf{y}\,|\,\mathbf{x})+\alpha\log P_{\text{LM}}(\mathbf{y})+\beta|\mathbf{y}|
\]
is identical in form, but $P_{\text{ASR}^{\ast}}$ now reflects the domain-adapted Whisper parameters.  All other decoding details (cleared \texttt{suppress\_tokens}, \texttt{max\_new\_tokens}\,$=180$, GPT-2 window limit) are inherited unchanged.
 
\textbf{MAML.} We meta-train \texttt{openai/whisper-base} with first-order MAML while keeping the backbone frozen and updating rank-8 LoRA adapters ($\alpha=16$, dropout $=$ 0.05) inserted into every Q, K, V, O projection of the encoder/decoder self-attention layers.
Each meta-batch contains two tasks—one drawn from the supported-language pool and one from the unsupported-language pool—and for every task, we sample $k=8$ support and $q=8$ query utterances ($\leq$ 15s).
A learner clone performs three inner-loop gradient steps on the support set with learning rate $1\times10^{-4}$; the resulting weights are evaluated on the query set, and the query loss from both tasks is back-propagated through the pre-adaptation adapter parameters.
These outer-loop gradients are updated using AdamW (learning rate $2\times10^{-4}$, 10\% cosine warm-up, gradient clip 1.0) for a total of 1,000 meta-updates ($\approx$10 meta-epochs given a 500-sample cap per language).
The trained initialization is then fast-adapted to each unseen target language by fine-tuning the adapters alone for 100 mini-batches (batch 2, learning rate $5 \times 10^{-5}$) before computing CER/WER, enabling a memory-efficient MAML baseline that fits comfortably on a single 80 GB H100 GPU.

\textbf{LoRA-Whisper (LoRA/AdaLoRA variant).}
We fine-tune \texttt{openai/whisper-large-v2} with LoRA and use the AdaLoRA variant that dynamically allocates rank during training.
The backbone weights are kept frozen, and only the LoRA parameters are updated during fine-tuning.
All hyperparameters are listed in Table~\ref{tab:hyperparameter}.

\begin{table*}
\centering
\normalsize
\caption[Hyperparameters]{\textbf{Hyperparameters.}
This table reports the hyperparameters used for all the methods. SICL and SMICL are meta-learning approaches using in-context learning, while Normal LoRA and Normal AdaLoRA are standard fine-tuning approaches with parameter-efficient training.}
\scalebox{0.95}{\begin{tabular}{@{}ccccccc@{}}\toprule
\textbf{Method} & \textbf{Hyperparameter} & 
\textbf{Value} \\
\cmidrule{1-3}
\multirow{8}{*}{SMICL} & Learning Rate & 1e-3 \\
& Batch Size & 4 \\
& \# Steps & 300 \\
& Warmup Steps & 100 \\
& PEFT Mode & AdaLoRA \\
& AdaLoRA Init R & 12 \\
& AdaLoRA Target R & 4 \\
\cmidrule{1-3}
\multirow{7}{*}{LoRA-Whisper (LoRA)} & Learning Rate & 5e-5 \\
& Batch Size & 4 \\
& \# Steps & 300 \\
& Warmup Steps & 100 \\
& PEFT Mode & LoRA \\
& LoRA R & 64 \\
& LoRA Alpha & 128 \\
& LoRA Dropout & 0.05 \\
\cmidrule{1-3}
\multirow{8}{*}{LoRA-Whisper (AdaLoRA)} & Learning Rate & 1e-3 \\
& Batch Size & 4 \\
& \# Steps & 300 \\
& Warmup Steps & 100 \\
& PEFT Mode & AdaLoRA \\
& AdaLoRA Init R & 12 \\
& AdaLoRA Target R & 4 \\
& AdaLoRA Alpha & 32 \\
& AdaLoRA Dropout & 0.1 \\
\bottomrule
\end{tabular}}
\label{tab:hyperparameter}
\end{table*} 

\begin{algorithm}[H]
\caption{Monte-Carlo Estimation of Typology Index ($\mathrm{TI}$) and Jaccard Coverage ($\mathrm{Jmm}$)}
\label{alg:mc_ti_jmm}
\textbf{Input}: Whisper embedding cache $\mathcal{C}$, subset sizes $K=\{k_1,\dots,k_L\}$, repetitions $B$, PCA dimension $d$ (default $128$), metric set $\mathcal{M}\subseteq\{\mathrm{TI},\mathrm{Jmm}\}$, random seed $s$ \\
\textbf{Output}: Estimates $\bigl\{(\mu_{k,m},\sigma_{k,m}) \mid k\!\in\!K,\; m\!\in\!\mathcal{M}\bigr\}$

\begin{algorithmic}[1]
\STATE {\color{gray}// Pre-processing}
\STATE Load utterance-level embeddings from $\mathcal{C}$ and \textbf{average} them per language \label{step:load}
\STATE Concatenate all language vectors: $E \leftarrow [\mathbf{e}_1;\dots;\mathbf{e}_N] \in\mathbb{R}^{N\times D_{\text{raw}}}$
\STATE Apply PCA: $\hat{E}\!=\!\text{PCA}_d(E)$,      \hfill {\scriptsize retains $>95\%$ variance}
\STATE Compute global quartile thresholds $\boldsymbol{\tau}^{(1:3)}$ over $\hat{E}$ \label{step:quartile}
\STATE Discretise each component into 4 one-hot bins (quartile coding) to form binary matrix $M\!\in\!\{0,1\}^{N\times4d}$

\STATE {\color{gray}// Monte-Carlo loop over subset sizes}
\FOR{each $k\in K$}  \label{step:kloop}
    \FOR{each $m\in\mathcal{M}$}  \label{step:metricloop}
        \STATE Initialise list $\mathrm{vals}\leftarrow[\;]$
        \FOR{$b=1$ \textbf{to} $B$}  \label{step:mc}
            \STATE Sample subset indices $S_b\subseteq\{1,\dots,N\},\;|S_b|=k$ with RNG$(s)$
            \STATE $M_b \leftarrow M_{S_b,:}$                                         \hfill {\scriptsize binary slice}
            \IF{$m = \mathrm{TI}$}
                \STATE $p \leftarrow \tfrac{1}{k}\sum_{\ell\in S_b} M_{\ell,:}$      \hfill {\scriptsize Bernoulli means}
                \STATE $v_b \leftarrow \frac{1}{4d}\sum_{i=1}^{4d}
                       \bigl[-p_i\log_2 p_i -(1-p_i)\log_2(1-p_i)\bigr]$
            \ELSE[{$m = \mathrm{Jmm}$}]
                \STATE $\operatorname{cover}_F \leftarrow \mathbf{1}[M^\top\mathbf{1}>0]$
                \STATE $\operatorname{cover}_S \leftarrow \mathbf{1}[M_b^\top\mathbf{1}>0]$
                \STATE $v_b \leftarrow
                  \dfrac{\lVert\operatorname{cover}_S \land \operatorname{cover}_F\rVert_1}
                        {\lVert\operatorname{cover}_S \lor  \operatorname{cover}_F\rVert_1}$ \label{step:jmm}
            \ENDIF
            \STATE Append $v_b$ to $\mathrm{vals}$
        \ENDFOR
        \STATE $\mu_{k,m} \leftarrow \text{mean}(\mathrm{vals}),\;
               \sigma_{k,m} \leftarrow \text{std}(\mathrm{vals},\text{ddof}=1)$
        \STATE \textbf{Report} $(\mu_{k,m},\sigma_{k,m})$
    \ENDFOR
\ENDFOR
\STATE {\color{gray}// Sanity-check (optional): ensure $\mathrm{Jmm}(M,M)=1$}
\IF{$\mathrm{Jmm}\in\mathcal{M}$}
    \STATE \textbf{assert} $\bigl|\mathrm{Jmm}(M,M)-1\bigr| < 10^{-6}$
\ENDIF
\STATE \textbf{return} $\bigl\{(\mu_{k,m},\sigma_{k,m})\bigr\}$
\end{algorithmic}
\end{algorithm}

\subsection{Detail Results on All Languages}

Table~\ref{tab:all_languages} presents the detailed Character Error Rates (CER) for all 134 languages. Although \texttt{org\_jpn} and \texttt{jpn} refer to the same language, they are reported separately for completeness.

\input{tables/all_languages}

\subsection{Detailed Error Analysis}

For each language \texttt{div}, \texttt{sin}, and \texttt{nor}, we inspected both the time-aligned hypotheses and the references to classify errors into the three categories below:
\begin{itemize}
    \item \textbf{Script Mis-identification (\textsc{Script})}\\
          The decoder writes in a different script.
    \item \textbf{Degeneration Loop (\textsc{Repeat})}\\
          A burst of high-probability tokens is repeated,
          inflating the hypothesis length by $3$–$6$ times.
    \item \textbf{Hallucinated Completion / Code-Switch (\textsc{Halluc.})}\\
          The model outputs a well-formed clause that is much longer than the ground truth, but does not repeat characters.
\end{itemize}

\begin{figure}
    \centering
    \includegraphics[width=\linewidth]{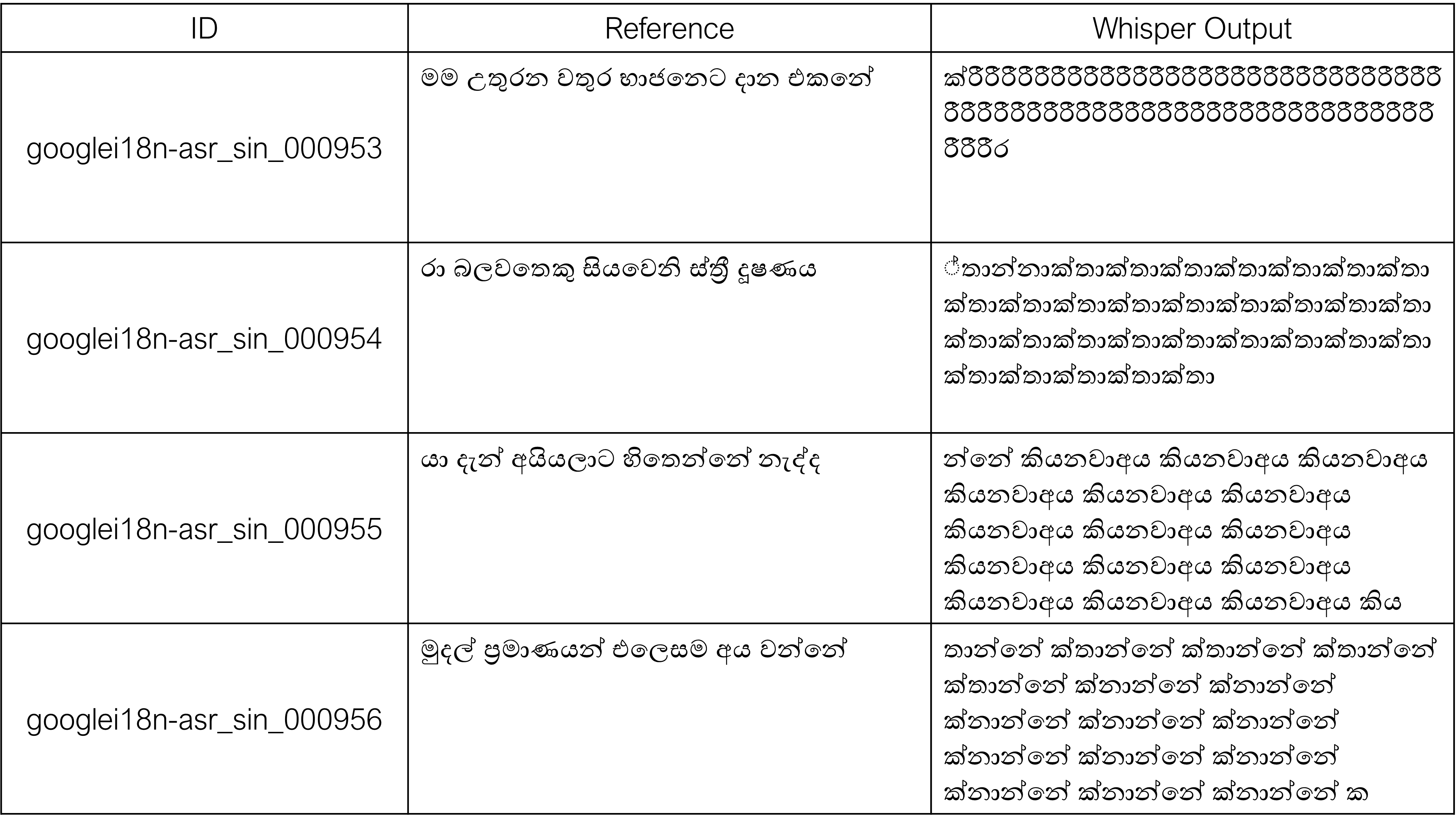}
    \caption{\textbf{Examples of the \textsc{Repeat} error in \texttt{sin}.}
For four utterances from the Google i18n-ASR test set, we list the utterance ID, the reference transcription, and Whisper’s hypothesis. In every case, the Whisper copies a short sequence of characters or syllables many times, producing an output that is vastly longer than the ground truth.}
    \label{fig:sin}
\end{figure}

\begin{figure}
    \centering
    \includegraphics[width=\linewidth]{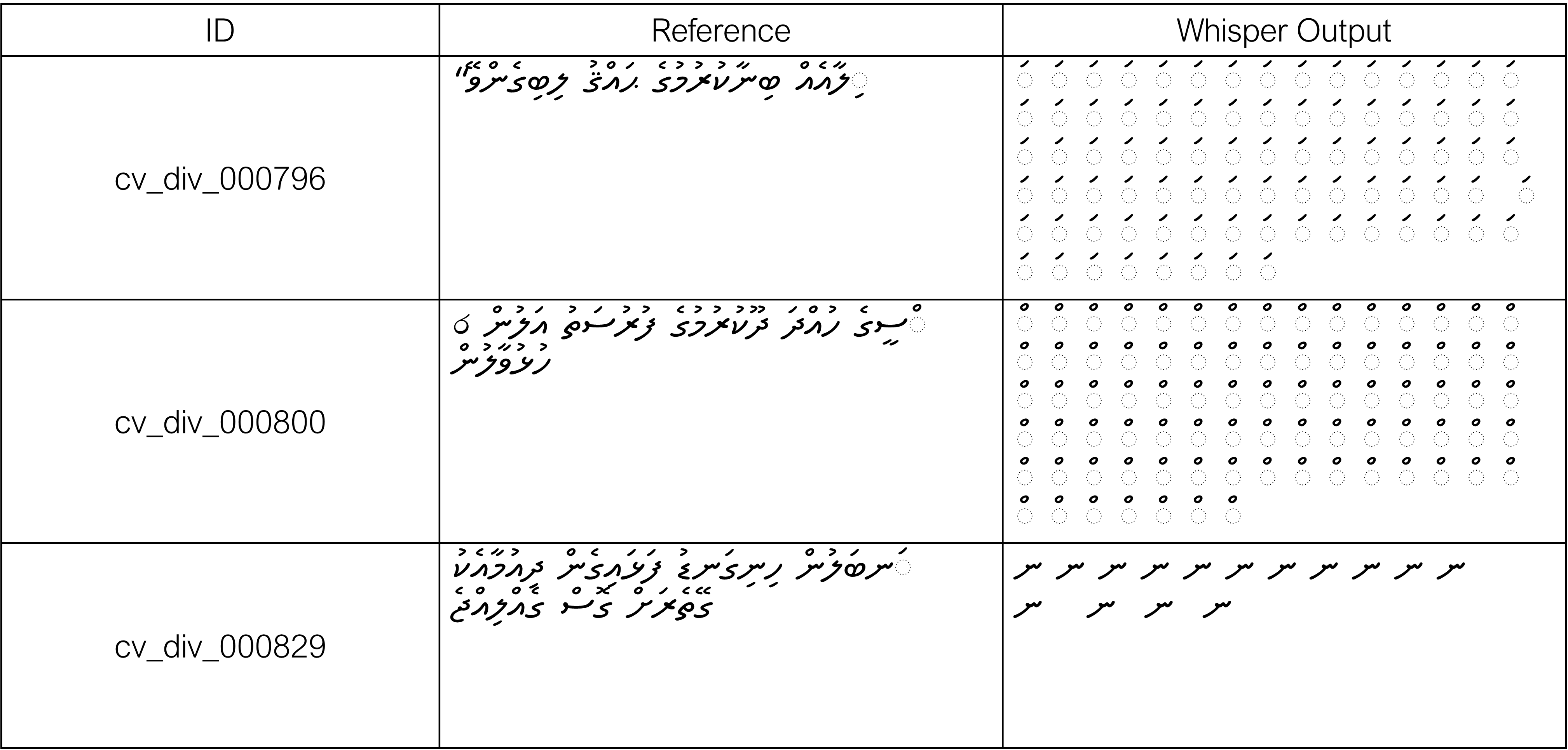}
    \caption{\textbf{Examples of the \textsc{Repeat} error in \texttt{div}.}  
Three utterances from the Common Voice test set are shown with their reference transcriptions and Whisper outputs.  
In every case, the model collapses into endlessly repeating a single character (or a very short substring of it), yielding an output that is orders of magnitude longer than the ground truth and carries no linguistic content.}

    \label{fig:div}
\end{figure}

For \texttt{sin} and \texttt{div}, the predominant error type is \textsc{Repeat}, as illustrated in Fig.~\ref {fig:sin} and Fig.~\ref{fig:div}. Whisper frequently decodes utterances in these languages with repeated segments.

\begin{figure}
    \centering
    \includegraphics[width=\linewidth]{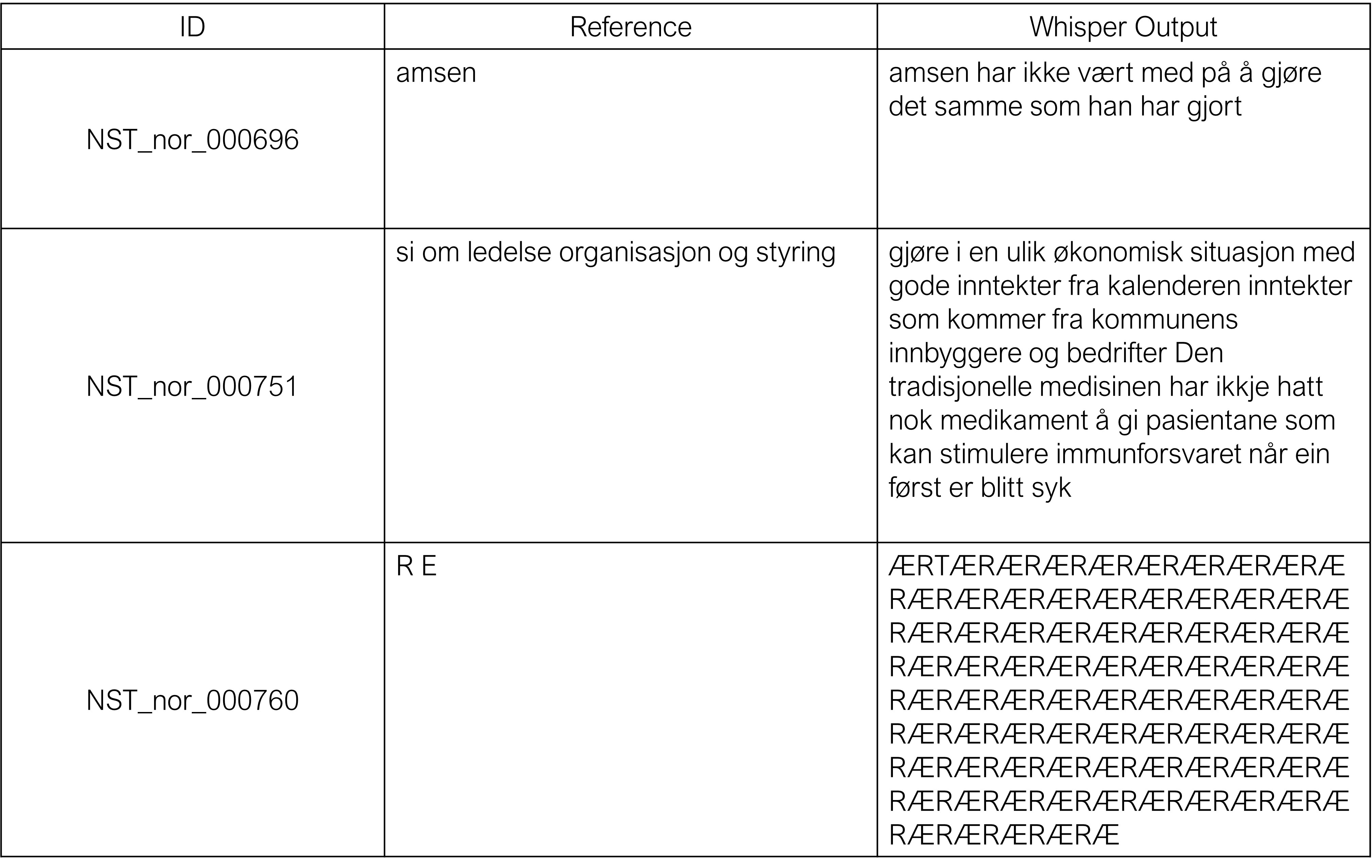}
    \caption{\textbf{Examples of \textsc{Script} and \textsc{Halluc.} errors in \texttt{nor}.}  
Rows are taken from the NST Norwegian corpus.  
Whisper sometimes invents fluent-sounding but spurious \texttt{nor} sentences that greatly exceed the length of the reference (first two rows, \textsc{Halluc.}), and at other times emits long strings of a single diacritic-rich character (\textsc{Script} error, bottom row).  Both phenomena result in transcripts that are largely unrelated to the intended utterance.}
    \label{fig:nor}
\end{figure}

In contrast, for \texttt{nor} the most common errors are \textsc{Script} and \textsc{Halluc.}. As shown in Fig.~\ref{fig:nor}, Whisper often produces transcriptions that are considerably longer than the ground truth, occasionally switching scripts or hallucinating additional content.

\subsection{Estimating Linguistic Diversity with TI and Jmm}
\label{sec:ti-jmm}

\textbf{Pre-processing.}
For each language~$\ell$, we average all Whisper encoder embeddings
extracted from its utterances, obtaining a single vector
$\mathbf{e}_\ell\!\in\!\mathbb{R}^{D_{\text{raw}}}$.
We then apply Principal Component Analysis (PCA) to reduce the
dimension to~$d$ (default $d\!=\!128$), preserving more than
$95\%$ of the variance in all our experiments.

\textbf{Quartile discretization.}
Let $\mathbf{z}_\ell\!\in\!\mathbb{R}^{d}$ be the PCA-projected vector.
For every component $j$ we compute its 25\%, 50\%, and 75\% percentiles
over \emph{all} languages, yielding the bin thresholds
$\boldsymbol{\tau}_j^{(1\!:\!3)}$.
Each $\mathbf{z}_\ell$ is then converted into a one-hot
\emph{4-level} code
\[
  \mathbf{m}_\ell \;=\;
  \bigl[\,
    \mathbf{1}[z_{\ell j}<\boldsymbol{\tau}^{(1)}_j],
    \mathbf{1}[\boldsymbol{\tau}^{(1)}_j\!\le z_{\ell j}<\boldsymbol{\tau}^{(2)}_j],
    \mathbf{1}[\boldsymbol{\tau}^{(2)}_j\!\le z_{\ell j}<\boldsymbol{\tau}^{(3)}_j],
    \mathbf{1}[z_{\ell j}\ge\boldsymbol{\tau}^{(3)}_j]
  \bigr]_{j=1}^{d}
  \;\in\;\{0,1\}^{4d}.
\]
Stacking all languages forms the binary matrix
$M\!\in\!\{0,1\}^{N\times 4d}$.

\textbf{Typology Index (TI).}
For a language subset $S$ of size $k$ we compute the empirical Bernoulli
means $p_i=\tfrac{1}{k}\sum_{\ell\in S}M_{\ell i}$ and the binary entropy
$H_2(p_i)= -p_i\log_2p_i-(1-p_i)\log_2(1-p_i)$.
The Typology Index is the average entropy over all bins:
\[
  \mathrm{TI}(S)=\frac{1}{4d}\sum_{i=1}^{4d} H_2\!\bigl(p_i\bigr).
\]
$\mathrm{TI}=1$ indicates a perfectly balanced presence/absence
distribution, i.e.\ maximal phonological diversity.

\textbf{Jaccard Coverage (Jmm).}
Given the full language set $F$ and its binary coverage vector
$\operatorname{cover}(F) = \mathbf{1}[M^\top\!\mathbf{1}>0]$, the coverage of a subset $S$ is
$\operatorname{cover}(S)=\mathbf{1}[M_S^\top\!\mathbf{1}>0]$.
Boolean-Jaccard coverage is
\[
  \mathrm{Jmm}(S,F)=
  \frac{\bigl\lVert\operatorname{cover}(S)\land\operatorname{cover}(F)\bigr\rVert_1}
       {\bigl\lVert\operatorname{cover}(S)\lor \operatorname{cover}(F)\bigr\rVert_1},
  \qquad
  0 \le \mathrm{Jmm}\le 1,
\]
where $\land/\lor$ and $\lVert\cdot\rVert_1$ denote element-wise \textsc{and}/\textsc{or} and~$\ell_1$ norm, respectively.

\textbf{Monte-Carlo estimation.}
Because the closed-form distribution of $\mathrm{TI}$ and $\mathrm{Jmm}$ over all $\binom{N}{k}$ subsets is intractable, we adopt a Monte-Carlo strategy (Algorithm~\ref{alg:mc_ti_jmm}).
For each subset size $k$ we draw~$B$ random subsets ($B=1000$ in our study) and report the sample mean~$\mu$ and unbiased standard deviation~$\sigma$ of the metric values.

%% file: tables/all_languages.tex
\begin{longtable}{c|>{\raggedleft\arraybackslash}p{2.2cm}|>{\raggedleft\arraybackslash}p{2.2cm}|>{\raggedleft\arraybackslash}p{2.2cm}|>{\raggedleft\arraybackslash}p{2.2cm}|>{\raggedleft\arraybackslash}p{2.2cm}|>{\raggedleft\arraybackslash}p{2.2cm}}

\caption{Character error rate (CER) for all languages, comparing SMILE-UL, SICL, LoRA-Whisper (LoRA), LoRA-Whisper (AdaLoRA), SMILE, and the unmodified Vanilla Whisper model. Lower percentages indicate more accurate transcriptions.}\label{tab:all_languages}\\
\toprule
\multicolumn{1}{c|}{Language} & \multicolumn{1}{c|}{SMILE-UL} & \multicolumn{1}{c|}{SICL} & \multicolumn{1}{c|}{\makecell{LoRA-Whisper\\(LoRA)}} & \multicolumn{1}{c|}{\makecell{LoRA-Whisper\\(AdaLoRA)}} & \multicolumn{1}{c|}{SMILE} & \multicolumn{1}{c}{Vanilla} \\
\midrule
\endhead
\texttt{afr} & 19.12\% & 16.00\% & 28.53\% & 27.95\% & 14.73\% & 22.32\% \\
\texttt{amh} & 206.40\% & 194.64\% & 136.24\% & 141.46\% & 190.76\% & 205.32\% \\
\texttt{ara} & 9.18\% & 7.75\% & 25.86\% & 22.45\% & 7.59\% & 10.81\% \\
\texttt{ast} & 11.00\% & 13.11\% & 12.72\% & 13.22\% & 12.93\% & 14.69\% \\
\texttt{aze} & 21.41\% & 20.07\% & 68.67\% & 57.11\% & 13.81\% & 6.02\% \\
\texttt{azz} & 85.12\% & 162.69\% & 214.04\% & 221.58\% & 70.84\% & 691.09\% \\
\texttt{bak} & 56.59\% & 39.40\% & 55.36\% & 56.18\% & 38.74\% & 92.18\% \\
\texttt{bas} & 35.13\% & 41.65\% & 58.04\% & 49.92\% & 40.83\% & 90.13\% \\
\texttt{bel} & 17.31\% & 12.40\% & 22.63\% & 23.22\% & 13.56\% & 16.88\% \\
\texttt{ben} & 34.99\% & 79.94\% & 103.97\% & 101.35\% & 53.14\% & 101.16\% \\
\texttt{bre} & 38.52\% & 40.31\% & 53.74\% & 54.39\% & 40.18\% & 68.61\% \\
\texttt{bul} & 4.63\% & 7.59\% & 6.98\% & 6.97\% & 6.30\% & 5.13\% \\
\texttt{cat} & 4.93\% & 4.81\% & 7.30\% & 6.83\% & 4.77\% & 5.34\% \\
\texttt{ceb} & 14.08\% & 15.35\% & 8.47\% & 8.43\% & 15.28\% & 14.06\% \\
\texttt{ces} & 6.17\% & 6.81\% & 15.97\% & 37.37\% & 5.26\% & 12.87\% \\
\texttt{chv} & 71.18\% & 51.39\% & 42.21\% & 54.60\% & 43.48\% & 113.25\% \\
\texttt{ckb} & 77.79\% & 181.12\% & 75.59\% & 69.58\% & 104.93\% & 93.82\% \\
\texttt{cmn} & 15.38\% & 15.42\% & 62.80\% & 35.44\% & 15.20\% & 39.32\% \\
\texttt{cnh} & 31.36\% & 34.70\% & 53.22\% & 45.33\% & 35.27\% & 85.03\% \\
\texttt{cym} & 15.23\% & 13.56\% & 22.65\% & 23.00\% & 13.69\% & 13.73\% \\
\texttt{dan} & 7.74\% & 7.52\% & 30.06\% & 30.69\% & 8.42\% & 21.72\% \\
\texttt{deu} & 5.03\% & 5.20\% & 6.72\% & 7.72\% & 4.77\% & 6.51\% \\
\texttt{div} & 381.82\% & 336.61\% & 103.85\% & 100.29\% & 383.95\% & 135.88\% \\
\texttt{ell} & 5.33\% & 5.02\% & 28.81\% & 21.63\% & 5.58\% & 6.04\% \\
\texttt{eng} & 5.25\% & 5.49\% & 6.02\% & 6.53\% & 5.16\% & 5.93\% \\
\texttt{epo} & 9.36\% & 10.34\% & 23.69\% & 22.19\% & 10.43\% & 39.70\% \\
\texttt{est} & 8.66\% & 7.72\% & 14.50\% & 13.74\% & 7.72\% & 8.37\% \\
\texttt{eus} & 15.02\% & 8.77\% & 30.95\% & 34.92\% & 9.47\% & 12.25\% \\
\texttt{fas} & 24.06\% & 16.78\% & 69.65\% & 61.41\% & 19.74\% & 20.18\% \\
\texttt{fil} & 20.76\% & 20.94\% & 5.20\% & 5.84\% & 21.96\% & 6.38\% \\
\texttt{fin} & 2.61\% & 2.41\% & 3.59\% & 3.48\% & 2.49\% & 3.16\% \\
\texttt{fra} & 5.75\% & 5.82\% & 9.55\% & 9.99\% & 5.34\% & 9.49\% \\
\texttt{frr} & 27.99\% & 29.18\% & 31.64\% & 32.79\% & 30.17\% & 43.13\% \\
\texttt{ful} & 42.46\% & 39.52\% & 24.82\% & 27.29\% & 32.20\% & 80.76\% \\
\texttt{glg} & 6.08\% & 5.43\% & 10.45\% & 9.86\% & 5.63\% & 6.10\% \\
\texttt{grn} & 36.35\% & 35.16\% & 42.63\% & 45.35\% & 37.82\% & 70.77\% \\
\texttt{guj} & 30.48\% & 44.04\% & 97.76\% & 94.79\% & 70.37\% & 88.77\% \\
\texttt{hau} & 34.89\% & 33.93\% & 41.97\% & 47.38\% & 33.35\% & 39.66\% \\
\texttt{heb} & 15.29\% & 14.19\% & 32.96\% & 31.23\% & 13.62\% & 13.32\% \\
\texttt{hin} & 13.07\% & 10.59\% & 93.67\% & 90.52\% & 11.56\% & 12.52\% \\
\texttt{hrv} & 7.64\% & 5.23\% & 26.54\% & 27.45\% & 4.28\% & 5.76\% \\
\texttt{hsb} & 22.63\% & 27.03\% & 50.69\% & 58.30\% & 27.03\% & 50.58\% \\
\texttt{hun} & 6.17\% & 6.09\% & 7.03\% & 6.69\% & 4.93\% & 5.21\% \\
\texttt{hye} & 18.36\% & 13.72\% & 93.45\% & 93.10\% & 15.30\% & 24.89\% \\
\texttt{ibo} & 57.80\% & 51.56\% & 40.76\% & 40.07\% & 50.88\% & 70.81\% \\
\texttt{ina} & 6.58\% & 8.56\% & 10.67\% & 11.39\% & 8.94\% & 24.23\% \\
\texttt{ind} & 3.47\% & 3.40\% & 6.03\% & 5.89\% & 3.35\% & 4.59\% \\
\texttt{isl} & 21.99\% & 16.19\% & 19.83\% & 19.39\% & 17.15\% & 12.60\% \\
\texttt{ita} & 6.07\% & 4.02\% & 5.11\% & 4.95\% & 3.90\% & 4.98\% \\
\texttt{jav} & 15.32\% & 14.73\% & 18.88\% & 18.04\% & 15.02\% & 21.41\% \\
\texttt{jpn} & 8.36\% & 7.62\% & 21.82\% & 23.71\% & 7.91\% & 17.14\% \\
\texttt{kab} & 84.44\% & 487.24\% & 134.24\% & 118.45\% & 84.70\% & 97.37\% \\
\texttt{kam} & 44.18\% & 46.49\% & 32.80\% & 33.26\% & 47.82\% & 144.92\% \\
\texttt{kan} & 20.39\% & 26.06\% & 107.95\% & 102.36\% & 43.38\% & 74.31\% \\
\texttt{kat} & 1256.91\% & 2049.49\% & 191.42\% & 121.02\% & 1741.69\% & 397.54\% \\
\texttt{kaz} & 17.26\% & 13.18\% & 25.91\% & 25.55\% & 13.02\% & 13.67\% \\
\texttt{kea} & 21.53\% & 28.96\% & 21.67\% & 24.10\% & 27.04\% & 56.01\% \\
\texttt{khm} & 140.76\% & 114.53\% & 113.32\% & 111.59\% & 129.37\% & 130.16\% \\
\texttt{kin} & 44.80\% & 60.55\% & 73.57\% & 93.20\% & 55.48\% & 140.40\% \\
\texttt{kir} & 31.40\% & 56.70\% & 60.07\% & 57.32\% & 45.82\% & 100.26\% \\
\texttt{kmr} & 38.50\% & 40.29\% & 80.82\% & 73.70\% & 41.78\% & 75.04\% \\
\texttt{kor} & 33.45\% & 22.52\% & 33.81\% & 33.83\% & 30.25\% & 32.23\% \\
\texttt{lao} & 182.48\% & 155.30\% & 110.73\% & 112.06\% & 176.67\% & 99.93\% \\
\texttt{lav} & 7.77\% & 5.66\% & 48.74\% & 51.42\% & 5.54\% & 7.75\% \\
\texttt{lga} & 31.88\% & 34.71\% & 35.85\% & 37.26\% & 35.16\% & 83.98\% \\
\texttt{lin} & 43.01\% & 44.83\% & 22.47\% & 22.45\% & 43.37\% & 18.74\% \\
\texttt{lit} & 13.65\% & 9.76\% & 52.63\% & 55.15\% & 11.23\% & 11.11\% \\
\texttt{ltz} & 29.06\% & 57.74\% & 35.80\% & 34.23\% & 28.17\% & 33.24\% \\
\texttt{lug} & 34.88\% & 37.62\% & 25.55\% & 25.63\% & 36.31\% & 145.14\% \\
\texttt{luo} & 32.05\% & 33.66\% & 24.73\% & 25.06\% & 31.26\% & 119.84\% \\
\texttt{mal} & 45.76\% & 73.70\% & 106.29\% & 99.69\% & 66.52\% & 104.85\% \\
\texttt{mar} & 19.97\% & 19.14\% & 91.08\% & 93.66\% & 20.26\% & 24.00\% \\
\texttt{mhr} & 22.74\% & 41.45\% & 32.20\% & 29.18\% & 44.07\% & 85.80\% \\
\texttt{mkd} & 5.26\% & 4.24\% & 7.84\% & 7.39\% & 4.73\% & 4.48\% \\
\texttt{mlt} & 29.10\% & 31.91\% & 39.95\% & 47.44\% & 33.91\% & 41.56\% \\
\texttt{mon} & 58.35\% & 78.26\% & 80.41\% & 79.21\% & 69.39\% & 102.05\% \\
\texttt{mri} & 42.22\% & 41.42\% & 12.66\% & 11.83\% & 41.77\% & 9.12\% \\
\texttt{mrj} & 26.02\% & 67.79\% & 49.51\% & 58.17\% & 58.28\% & 87.14\% \\
\texttt{msa} & 6.52\% & 35.82\% & 4.49\% & 4.33\% & 6.18\% & 4.24\% \\
\texttt{mya} & N/A & N/A & 96.48\% & 97.89\% & N/A & 103.99\% \\
\texttt{myv} & 26.33\% & 56.61\% & 61.09\% & 53.80\% & 49.01\% & 101.05\% \\
\texttt{nbl} & 20.64\% & 73.93\% & 22.65\% & 22.83\% & 35.99\% & 108.58\% \\
\texttt{nep} & 39.18\% & 66.25\% & 106.20\% & 108.94\% & 42.82\% & 38.06\% \\
\texttt{nld} & 5.34\% & 5.37\% & 8.92\% & 9.67\% & 4.94\% & 9.08\% \\
\texttt{nno} & 10.06\% & 8.40\% & 21.82\% & 21.98\% & 8.87\% & 19.76\% \\
\texttt{nor} & 555.56\% & 814.47\% & 279.70\% & 281.02\% & 497.73\% & 282.26\% \\
\texttt{nso} & 35.74\% & 40.30\% & 32.42\% & 31.78\% & 64.92\% & 97.03\% \\
\texttt{nya} & 37.68\% & 42.77\% & 26.49\% & 27.19\% & 42.32\% & 90.48\% \\
\texttt{oci} & 29.49\% & 28.97\% & 26.99\% & 27.20\% & 27.55\% & 24.62\% \\
\texttt{org\_jpn} & 7.70\% & 6.98\% & 24.23\% & 23.71\% & 7.94\% & 19.66\% \\
\texttt{orm} & 32.23\% & 48.17\% & 30.34\% & 34.23\% & 34.51\% & 68.92\% \\
\texttt{ory} & 47.03\% & 102.35\% & 108.26\% & 104.65\% & 119.91\% & 104.40\% \\
\texttt{pan} & 44.32\% & 35.87\% & 91.59\% & 90.05\% & 47.41\% & 87.08\% \\
\texttt{pol} & 3.80\% & 3.45\% & 3.70\% & 3.71\% & 3.32\% & 2.95\% \\
\texttt{por} & 3.65\% & 14.43\% & 8.23\% & 7.90\% & 6.37\% & 4.75\% \\
\texttt{pus} & 42.13\% & 37.85\% & 52.49\% & 61.27\% & 38.59\% & 60.40\% \\
\texttt{ron} & 4.66\% & 3.95\% & 8.45\% & 7.86\% & 3.73\% & 4.34\% \\
\texttt{rus} & 3.27\% & 6.28\% & 4.00\% & 4.06\% & 3.18\% & 4.82\% \\
\texttt{sah} & 32.65\% & 56.30\% & 42.07\% & 40.36\% & 50.52\% & 108.20\% \\
\texttt{sin} & 644.02\% & 582.87\% & 109.17\% & 109.91\% & 549.32\% & 535.45\% \\
\texttt{skr} & 54.96\% & 90.33\% & 84.29\% & 91.81\% & 75.00\% & 97.05\% \\
\texttt{slk} & 6.91\% & 5.33\% & 19.07\% & 18.62\% & 5.78\% & 10.22\% \\
\texttt{slv} & 10.70\% & 9.30\% & 34.76\% & 43.76\% & 8.50\% & 9.85\% \\
\texttt{sna} & 30.87\% & 30.43\% & 29.20\% & 29.38\% & 31.83\% & 83.74\% \\
\texttt{snd} & 55.26\% & 53.13\% & 65.43\% & 59.32\% & 50.31\% & 75.85\% \\
\texttt{som} & 38.28\% & 40.21\% & 53.98\% & 56.18\% & 42.79\% & 128.51\% \\
\texttt{sot} & 22.58\% & 52.54\% & 24.03\% & 26.02\% & 28.19\% & 76.02\% \\
\texttt{spa} & 3.13\% & 3.33\% & 3.34\% & 3.32\% & 2.56\% & 3.16\% \\
\texttt{srp} & 11.28\% & 9.97\% & 39.25\% & 36.46\% & 10.02\% & 39.68\% \\
\texttt{ssw} & 21.69\% & 26.64\% & 28.28\% & 27.26\% & 39.67\% & 98.31\% \\
\texttt{sun} & 6.53\% & 6.66\% & 19.46\% & 19.41\% & 6.68\% & 23.82\% \\
\texttt{swa} & 19.89\% & 32.60\% & 21.96\% & 21.70\% & 23.98\% & 58.31\% \\
\texttt{swe} & 6.43\% & 5.14\% & 12.81\% & 11.37\% & 5.26\% & 9.13\% \\
\texttt{tam} & 18.83\% & 35.09\% & 90.67\% & 86.95\% & 22.46\% & 56.84\% \\
\texttt{tat} & 30.96\% & 31.22\% & 52.38\% & 46.61\% & 38.43\% & 87.10\% \\
\texttt{tel} & 26.25\% & 48.74\% & 96.01\% & 97.44\% & 36.65\% & 88.95\% \\
\texttt{tgk} & 26.54\% & 39.67\% & 44.78\% & 47.69\% & 28.78\% & 37.14\% \\
\texttt{tha} & 13.40\% & 11.27\% & 30.63\% & 35.44\% & 12.49\% & 11.84\% \\
\texttt{tok} & 6.19\% & 5.25\% & 12.20\% & 12.31\% & 5.76\% & 32.24\% \\
\texttt{tos} & 77.04\% & 152.58\% & 56.64\% & 63.03\% & 105.19\% & 136.53\% \\
\texttt{tsn} & 22.64\% & 35.57\% & 26.33\% & 26.72\% & 32.07\% & 92.65\% \\
\texttt{tso} & 21.66\% & 27.35\% & 28.95\% & 25.11\% & 27.77\% & 105.11\% \\
\texttt{tur} & 7.53\% & 8.27\% & 26.74\% & 15.15\% & 7.06\% & 8.37\% \\
\texttt{uig} & 79.58\% & 202.10\% & 89.11\% & 84.19\% & 140.53\% & 98.71\% \\
\texttt{ukr} & 5.51\% & 4.31\% & 9.08\% & 9.04\% & 4.33\% & 5.47\% \\
\texttt{umb} & 44.61\% & 43.87\% & 19.82\% & 24.90\% & 43.84\% & 189.58\% \\
\texttt{urd} & 9.24\% & 7.06\% & 23.77\% & 21.50\% & 8.26\% & 9.79\% \\
\texttt{uzb} & 26.24\% & 22.13\% & 99.77\% & 97.80\% & 22.74\% & 30.05\% \\
\texttt{ven} & 30.51\% & 66.18\% & 30.18\% & 30.14\% & 54.46\% & 70.19\% \\
\texttt{vie} & 9.21\% & 9.40\% & 14.90\% & 14.33\% & 7.48\% & 11.30\% \\
\texttt{wol} & 31.43\% & 34.59\% & 31.23\% & 31.66\% & 35.99\% & 157.36\% \\
\texttt{xho} & 23.62\% & 28.11\% & 24.34\% & 25.67\% & 32.45\% & 127.78\% \\
\texttt{xty} & 87.13\% & 187.13\% & 72.12\% & 94.81\% & 132.42\% & 102.09\% \\
\texttt{yor} & 85.24\% & 80.22\% & 61.65\% & 61.76\% & 85.22\% & 66.30\% \\
\texttt{yue} & 11.07\% & 10.65\% & 24.53\% & 22.86\% & 11.48\% & 352.95\% \\
\bottomrule
\end{longtable}